\documentclass{aa}  
\usepackage{graphicx}
\usepackage{epstopdf}
\usepackage{amsmath}
\usepackage{txfonts}
\usepackage[usenames, dvipsnames]{xcolor}
\usepackage{textcomp,gensymb}
\usepackage[normalem]{ulem}
\usepackage{natbib}
\usepackage{tabu}
\begin{document} 

   \title{Dynamical ages of the Young Local Associations with \textit{Gaia}}
\titlerunning{Dynamical Ages of YLAs}

   \author{N. Miret-Roig\inst{1,2}, T. Antoja\inst{1}, M. Romero-G\'{o}mez\inst{1}  \& F. Figueras\inst{1} } 
         
   \institute{Institut de Ci\`{e}ncies del Cosmos, Universitat  de  Barcelona  (IEEC-UB), Mart\'{i} i Franqu\`{e}s  1, E-08028 Barcelona, Spain,  \email{nuria.miret-roig@u-bordeaux.fr}
   \and
              Laboratoire d'astrophysique de Bordeaux, Univ. Bordeaux, CNRS, B18N, allée Geoffroy Saint-Hilaire, 33615 Pessac, France  }

%   \date{Received September 15, 1996; accepted March 16, 1997}

% \abstract{}{}{}{}{} 
% 5 {} token are mandatory
  
  \abstract
   {The Young Local Associations (YLAs) constitute an excellent sample for the  study of a variety of astrophysical topics, especially the star formation process in low-density environments. Data  from the \textit{Gaia} mission allows us to undertake studies of the YLAs with unprecedented accuracy.}
   {We  determine the dynamical age and place of birth of a set of associations in a uniform and dynamically consistent manner. There are nine  YLAs in our sample $\epsilon$~Chamaeleontis, TW~Hydrae, $\beta$~Pictoris, Octans, Tucana-Horologium, Columba, Carina, Argus, and AB Doradus.}
   {We designed a method for  deriving the dynamical age of the YLAs based on the orbital integration. The method involves a strategy to account for the effect of  observational errors. We tested the method using mock YLAs. Finally, we applied it to our set of nine YLAs with astrometry from the first \textit{Gaia} data release and complementary on-ground radial velocities from the literature.}
   {Our orbital analysis yields a first estimate of the dynamical age of   3$^{+9}_{-0}$~Myr, 13$^{+7}_{-0}$~Myr, and 5$^{+23}_{-0}$~Myr for $\epsilon$~Chamaeleontis, $\beta$~Pictoris, and Tucana-Horologium, respectively. For four other associations (Octans, Columba, Carina, and Argus), we provide a lower limit for the dynamical age.  Our rigorous error treatment indicates that TW~Hydrae and AB~Doradus deserve further study. }
   {The dynamical ages that we obtain are compatible spectroscopic and isochrone fitting ages obtained elsewhere. From the orbital analysis, we suggest a scenario for these YLAs where there were two episodes of star formation: one $\sim40$~Myr ago in the first quadrant that gave birth to $\epsilon$~Chamaeleontis, TW~Hydrae, and $\beta$~Pictoris, and another $5-15$~Myr ago close to the Sun that formed Tucana-Horologium, Columba, and Carina. Future \textit{Gaia} data will provide the necessary accuracy to improve the present results, especially for the controversial age determinations, and additional evidence for the proposed scenario once a complete census of YLAs and better membership can be obtained.}


\maketitle
%
%________________________________________________________________

%\tableofcontents

%\newpage

\section{Introduction}\label{sec:introduction}

A Young Local Association (YLA) is a group of young stars  near the Sun; these stars are  thought to have formed together in the same star burst \citep{Jayawardhana00,deZeeuw99}. In consequence, the YLA members share common motions and are chemically homogeneous. With ages in the range from a few to several tens of Myr, YLAs are excellent tracers of the young population in the solar neighbourhood. They allow us to study the mechanisms driving the star formation process and also the secular evolution of the Milky Way thin disc. 

The first YLAs were discovered early in the 1990s through a combination of X-ray and optical spectroscopy  and kinematic data \citep[e.g.][]{delaReza89}. Thanks to the \textit{Hipparcos} astrometry, several new YLAs were found \citep[e.g.][]{Mamajek99,Zuckerman00}, and their membership and kinematic properties re-evaluated \citep[e.g.][]{Zuckerman04,Fernandez08,Barrado99,Barrado06}. In the last decade, several new surveys, still in progress, have aimed to search for nearby young associations and new low-mass members. Some examples are the SACY high-resolution optical spectroscopic survey \citep{Torres06} and the kinematically unbiased search combining the SuperWASP and the ROSAT all-sky surveys \citep{Binks15}. In addition, new codes have been developed to search for new low-mass star candidates such as BANYAN \citep{Malo13} and LACEwING \citep{Riedel17}. Despite this huge effort, future research still requires  a homogeneous, complete, and all-sky astrometric and spectroscopic survey, and this is a data feed that \textit{Gaia} is currently providing.

The first models to study the evolution of a YLA were developed in the 1960s. They were based on the notion of linear expansion, and considered that no forces  acted after birth \citep{Blaauw64}. Later on,  trace-back orbital analysis was undertaken and considered simplified Galactic potentials described by the Galactic epicycle theory (see \citealt{Makarov04} for a deep evaluation of this approximation). \citet{Brown97} combined N-body simulations with the effects of the Galactic tidal field from the epicyclic approximation; they   concluded that for associations with a velocity dispersion comparable to that of the YLAs (few km/s) the interaction between particles was unimportant. Others felt that the nearby interactions were only relevant at the very beginning of the initial expansion phases and that stars become unbound soon after gas expulsion from stellar winds \citep{Kroupa06}. Now  it is believed that the vast majority of clusters become unbound systems after birth \citep{Lada}. These are, therefore, the underlying assumptions of recent studies and we too use them  in the present work. 

The trace-back strategy  sheds light on the scenario for the star formation processes that took place in the solar neighbourhood in the last 50--100 Myr. \citet{Fernandez08}, from a compilation of \textit{Hipparcos} data, studied the kinematic evolution of YLAs and their relation to other young stellar complexes in the local interstellar medium. These authors used a Galactic potential with an axisymmetric component, and spiral arms and bar components. However, the large errors in the astrometry and the low number of members with available radial velocities did not allow them to accurately derive  the place of birth of these associations or their dynamical age. Other efforts came to similar conclusions. An example is the trace-back analysis carried out by \citet{Mamajek14} for the $\beta$ Pictoris YLA that, even considering a more sophisticated dynamical analysis with the NEMO stellar code \citep{Teuben95}, did not provide  clear evidence of expansion nor a clear trace-back age for this association. More recently, new strategies have been proposed such as the one of \citet{Riedel17} that presents membership studies combining all the kinematic and spatial information available with the results from the trace-back analysis (TRACEwING). Currently, this code uses only the epicyclic approximation for the orbit computation.

In the near future, the \textit{Gaia} mission \citep{GaiaColPrusti16} will provide us with three key contributions to the trace-back analysis of the YLAs: a realistic and accurate Galactic potential, an unprecedented accurate astrometric data for sources up to faint magnitudes $(G=21)$ corresponding to a large number of low-mass members of the YLAs, and a homogeneous and complete sample of newly discovered YLAs. The \textit{Gaia} era has just started with the first data release in September 2016. This release includes the Tycho-Gaia astrometric solution (TGAS) with parallax and proper motions with submilliarcsecond accuracy for a subset of $2.5$ million Tycho-2 stars.  
 
In this work we have used this new \textit{Gaia} data to trace-back the members of the known YLAs recently compiled by \citet{Riedel17}. We trace-back the orbits of the YLA members using a realistic 3D Galactic potential, and we determine the ages of the YLAs as the time when their members present the minimum dispersion in positions. We use the term `dynamical age' to denote this age determination\footnote{The term `kinematical ages' is sometimes used for similar purposes.}$^{,}$\footnote{This should not  be confused with the definition of the dynamical age as the number of relaxation times a given cluster has lived through \citep[see e.g.][]{Geller15}.}. The trace-back orbital study provides us with an age determination for the YLAs that is independent of the spectroscopic and isochronal ages.

This study is structured as follows. In Section~\ref{sec:sample} we present and characterise the observational data of the set of YLAs we study. In Section~\ref{sec:method-sim} we introduce our methodology to determine dynamical ages based on the orbital trace-back integration. We also describe  our approach to model the observational errors, and we test the method with simulated YLAs. In Section~\ref{sec:results} we show our results on the dynamical ages of the YLAs obtained and the study of their place of birth. In Section~\ref{sec:discussion} we discuss the results and present some future perspectives. Finally, we conclude in Sect.~\ref{sec:conclusions}.

%__________________________________________________________________

\section{YLAs data}
\label{sec:sample}

\begin{table*}
\centering
\tabcolsep=0.15cm
\begin{tabular}{llccccccccccc}
\hline
\hline\\[0.2mm]
\multicolumn{1}{l}{Name}  & \multicolumn{1}{l}{Abbreviation} & \multicolumn{4}{c}{Num. of Members} & $\bar{r}$ & $\bar{\xi^\prime}$ & $\bar{\eta^\prime}$ & $\bar{\zeta^\prime}$ & $\bar{U}$ & $\bar{V}$ & $\bar{W}$ \\
   & & CSNYS & TGAS & 6D & $2\sigma$ & (pc)  & (pc)  & (pc)  & (pc)  & (km s$^{-1}$)   & (km s$^{-1}$)   & (km s$^{-1}$) \\[1mm]
\hline\\[0.2mm]
$\epsilon$ Chamaeleontis & $\epsilon$ Cha & 43  & 13 & 10 & 10 & 101$\pm$2 & -49 & -85 & -22 & -10.1 & -19.3 & -9.7 \\
TW Hydrae                & TW Hya         & 59  & 7  & 7  & 7  & 66$\pm$8  & -16 & -60 &  20 & -10.9 & -20.2 & -5.6 \\
$\beta$ Pictoris         & $\beta$ Pic    & 146 & 32 & 31 & 21 & 40$\pm$4  & -17 &  -4 & -15 &  -9.9 & -16.2 & -9.0 \\
Octans                   & Oct            & 49  & 14 & 14 & 13 & 135$\pm$7 &   9 & -100 & -61 & -13.9 &  -3.6 & -10.3\\
Tucana-Horologium        & Tuc-Hor        & 250 & 52 & 49 & 30 & 47$\pm$1  &   0 & -22 & -35 &  -9.8 & -20.9 & -1.1 \\
Columba                  & Col            & 98  & 46 & 36 & 24 & 73$\pm$5  &  29 & -48 & -37 & -12.8 & -21.8 & -5.5 \\
Carina                   & Car            & 40  & 22 & 22 & 17 & 98$\pm$9  &  -8 & -94 & -17 & -10.3 & -22.8 & -4.6 \\
Argus                    & Arg            & 142 & 46 & 44 & 25 & 125$\pm$6 & -12 & -120 & -16 & -22.6 & -14.1 & -5.3 \\
AB Doradus               & AB Dor         & 207 & 69 & 61 & 38 & 48$\pm$5  &   5 & -14 & -20 &  -6.7 & -27.5 & -14.0\\[1mm]
\hline
\hline
\end{tabular}
\caption{Information on the YLAs used in our study. Columns indicate: 1) number of stars
catalogued as members in the CSNYS \citep{Riedel17}; 2) number of members with TGAS data; 3) number of stars with full 6D information; 4) number of remaining members after the $2\sigma$ cut in velocity space, which is the final set used to compute mean distances and heliocentric spatial and velocity components (Cols. 5 to 11). We compute the mean distance $\bar{r}$ (Col. 5) by inverting the parallax and its uncertainty as $\sigma_r/\sqrt{N}$. }
\label{tab:YLAs-review}
\end{table*}

Our methodology for deriving dynamical ages requires an initial sample of stars catalogued as members of a YLA. In addition, it is mandatory to have 6D phase space data for the present positions and velocities of these stars. As input catalogue we used the Catalog of Suspected Nearby Young Stars (CSNYS) recently published by \citet{Riedel17}. This catalogue contains $5350$ stars and was constructed from a wide variety of source papers which reported young  members of different nearby young moving groups and open clusters. From this sample we selected the associations with a reported maximum age of 150 Myr (see their Table~1). We  selected the individual members of the YLAs from those classified as members using the column `GROUP' of the CSNYS. As a result of this selection we obtained a list of $1034$ nearby stars belonging to nine YLAs\footnote{We do not consider the $\eta$ Cha and 32 Ori associations because they do not have  enough members after the cross-match with TGAS (see text).} (see Table~\ref{tab:YLAs-review}). Next, we cross-matched this catalogue with the \textit{Gaia} TGAS catalogue \citep{Lindegren16, Michalik15} using the Tycho-2 and Hipparcos IDs. As a result, we had a list of 301 stars with TGAS astrometry (30\% of the sample). For the radial velocities we used the CSNYS column `HRV' compiled from the literature, but we also did a cross-match with the RAdial Velocity Experiment \citep[RAVE,][]{Kunder17}. For stars having radial velocity from both sources, we took the one with the smaller reported error. Our final working sample had 274 stars from nine YLAs for which we were able to compute the 6D phase space coordinates\footnote{We used the initial compilation by \citet{Riedel17}, but not their outputs from the trace-back analysis or from their LACEwING code.}.

To place the stars in the configuration space we used the curvilinear heliocentric coordinates ($\xi^\prime$, $\eta^\prime$, $\zeta^\prime$) defined in \citet{Asiain99}. This coordinate system is centred on the Sun's current position, which we take as $R_\odot = 8.5$~kpc,  and rotates around the Galactic centre with a frequency of the circular velocity of $\omega_\odot = 25.88$~km~s$^{-1}$~kpc$^{-1}$. The radial component $\xi^\prime$ points to the Galactic anti-centre, the azimuthal component $\eta^\prime$ is measured along the circle of radius $R_\odot$ and is positive in the sense of the galactic rotation, and the vertical component  $\zeta^\prime$ is defined positive towards the north Galactic pole. This coordinate system minimises the variation in each component of the configuration space. For the velocity space we use the heliocentric Cartesian $(U, V, W)$ system, where $U$ is the velocity component positive towards the Galactic centre, $V$ is positive in the direction of the Galactic rotation, and $W$ is positive towards the north Galactic pole. For the coordinate transformation we use a peculiar solar motion of $(U_\odot, V_\odot, W_\odot) = (11.1,12.24,7.25)$~km~s$^{-1}$ \citep{Schonrich10}.

To define a first list of bona fide  members, we  traced back the orbits of all the stars with full 6D information using the axisymmetric potential (see Sect.~\ref{sec:method-sim}). The orbits of a few members depart significantly from the rest of orbits of the group. Therefore, we  applied a $2\sigma$ clipping in the curvilinear velocity space ($\dot{\xi^\prime},\dot{\eta^\prime},\dot{\zeta^\prime}$) in the present time. We did  this only for the associations with a number of members with full 6D data larger than 10, which excludes the $\epsilon$~Cha and TW Hya associations. The resultant bona fide sample contains 185 stars.

In Table~\ref{tab:YLAs-review} we present some information on the YLAs that we used in our analysis, including the number of members. Our sample contains fewer members than previous studies \citep[e.g.][]{Riedel17,Fernandez08} because we limited ourselves  to TGAS members in order to have homogeneous and high-quality astrometric data. The mean velocities of the YLAs (last columns) are similar to those of previous studies (e.g. \citealt{Fernandez08,Riedel17}). However, the mean positions of several associations differ from the values obtained in other studies, due to the selection cuts applied. The main characteristic of these associations is that they are concentrated in the velocity space. On the contrary, they are largely spread in positions, and therefore  a different selection of members leads to different mean positions.

\begin{figure}
\centering
\includegraphics[scale=0.35]{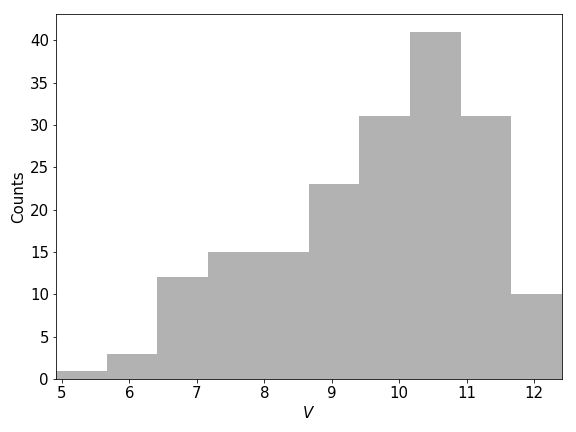}
\includegraphics[scale=0.35]{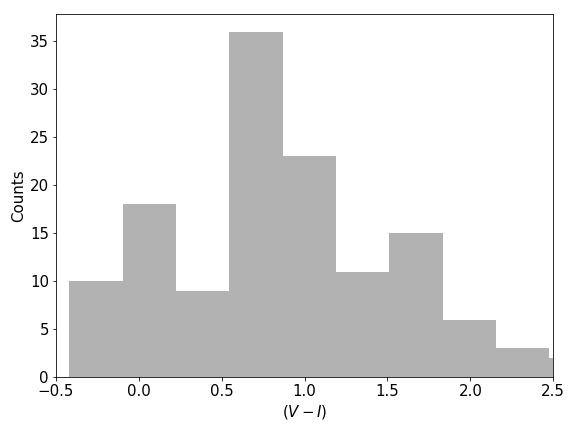}
\caption{Visual magnitude ($V$) distribution (top) and ($V-I$) colour distribution (bottom) of the bona fide members of the nine YLAs. For a few stars (28\% of the bona fide sample of 185 stars) the $V-I$ colour was not available.}
\label{fig:sample-distribution}
\end{figure}

In Figure~\ref{fig:sample-distribution} we present the distribution of visual apparent magnitude $V$ and colour $(V-I)$ for the 185 bona fide stars. The mean magnitude of the whole sample is $V=9.7$, which corresponds to $G=9.2$. The median TGAS error in parallax is $0.3$~mas and the median TGAS error in proper motion is $\sim 0.6$~mas~yr$^{-1}$. Since the parallax relative error is less than $2\%$ on average, we assume that the parallax inversion is an acceptable distance estimator. This gives a mean distance $\bar{r}$ of 75~pc taking all bona fide members of the nine YLAs. The resulting mean error in tangential velocity is $\sim 0.5$~km~s$^{-1}$. As mentioned above, radial velocities come from different sources and the estimated uncertainties provided by the authors range from 0.1 to 5~km~s$^{-1}$.

%__________________________________________________________________

\section{Methodology and tests with simulations}
\label{sec:method-sim}

In this section we describe our methodology and the  simulations used to test the capabilities of the trace-back orbital analysis to derive dynamical ages of the YLAs. First, we introduce the strategy to compute the stellar orbits and the Galactic potential that we used. Second, we explain how we generated simulated YLAs and their evolution. Then, we investigate a simple trace-back analysis of the simulated YLAs, without considering the observational errors. We also present a strategy to treat the effect of the observational errors on the trace-back analysis and we quantify their impact on the determination of the dynamical age. All this work allows us to present the criteria used to determine the dynamical age of the association and its uncertainty. At the end of this section, we decsribe the resampling method used to test the effect of having a restricted number of members.

\subsection{Galactic potential}
\label{subsec:integration}

We determined the stellar orbits through the integration of the equations of motion. 
For this we used the 3D Milky Way axisymmetric potential given by \cite{Allen91}, which consists of a spherical central bulge, a disc, and a massive spherical halo. In this model the total mass of the Milky Way is  assumed to be $9\times10^{11}$ M$_\odot$. The solar radius and the frequency of the circular orbit at this radius are those assumed also in Sect.~\ref{sec:sample}. We do not consider a Galactic bar nor a spiral arms potential since their effects are negligible for YLA-like orbits (nearly circular) placed near the Sun (see Sect.~\ref{subsec:spiral-arms} for a more detailed study on the effect of the spiral arms).

\subsection{Generation of simulated YLAs}
\label{subsec:generation-YLA}

Here we explain how we performed the simulations of YLAs. First, we took certain values for the centroid (mean position and velocity) of the YLA in the present $(t=0 \ \textup{Myr})$ and its age $(\tau)$. We integrated the orbit of this centroid back in time until the assumed age. At this point, we generated $N$ particles centred at the position of the centroid with Gaussian isotropic dispersions in positions and velocities in the Cartesian Galactocentric system. Then, we integrated the orbit of each particle forward in time from the birth moment until the present. Next, we took into account the observational constraints, i.e. we add observational errors to each particle. To do this, we assumed that the errors are Gaussian in all the observables. We considered the astrometric TGAS-like errors \citep[from Table 1 of][]{Michalik15} and an error in radial velocity of $\sigma_\textup{RV} =2$~km~s$^{-1}$, similar to that of the current data. Since the TGAS-like errors depend on the apparent visual magnitude ($V$) of the stars, we  assumed that our simulated YLAs have a Gaussian distribution of magnitudes centred at $V=9.7$ with a dispersion of 1.5~mag, similar to the observations (see Sect.~\ref{sec:sample}). For simplification purposes, we did not consider correlations among astrometric parameters. Then, we integrated the orbits backwards from the present until well beyond the age of the YLA (two times the age). Finally, we can analysed the evolution of the dispersion in positions as a function of time and determined the time of minimum dispersion which we give for the dynamical age. 

We  considered two different centroids in the present: the mean velocities and positions of AB Dor and those from the Hyades open cluster, published in \citet{Fernandez08} and \citet{Riedel17}, respectively. The first leads to an association with an almost circular orbit (orbit-1) and the other has a slightly more eccentric orbit (orbit-2). We also explored different initial dispersions, namely $1$~pc \citep{Pfalzner16} and $15$~pc \citep{Blaauw91} in positions, and $2$ and $4$~km~s$^{-1}$ in velocities \citep{Brown97}. With these values we \textbf{defined} two extreme cases for the initial conditions: IC-1 refers to the 1~pc and $2$~km~s$^{-1}$ case and IC-2 to the 15~pc and $4$~km~s$^{-1}$ case. In most cases we assumed an age of $\tau = 50$~Myr, but we explored other ages as well. 

We used $500$ particles in all the simulated YLAs. However, for the oldest YLAs and especially after adding observational errors, some members appeared very distant from the Sun in the present time. It would be difficult to classify these farther members  as members of a YLA. Indeed, most of the currently known members of the YLAs are within a distance of only $\sim100$~pc (see Sect.~\ref{sec:sample}), while in our simulations stars can be observed much farther than that. To mimic the real case, we applied a cut in distance at $500$~pc from the Sun\footnote{We used this number instead of 100~pc to account for a possible improvement on the limiting distance when new members are discovered with future observations.}. As an example, we find that in the present, after having applied the observational errors and for a YLA of $20$~Myr with IC-1 and following orbit-1, all the members are within a radius of $500$~pc; for a YLA of $50$~Myr about $1\%$ of members are farther than $500$~pc, and for a YLA of $80$~Myr about $10\%$ are farther than this.

\subsection{Orbital analysis of simulated YLAs}
\label{subsec:simple-case}

\begin{figure}
\centering
\includegraphics[scale=0.35]{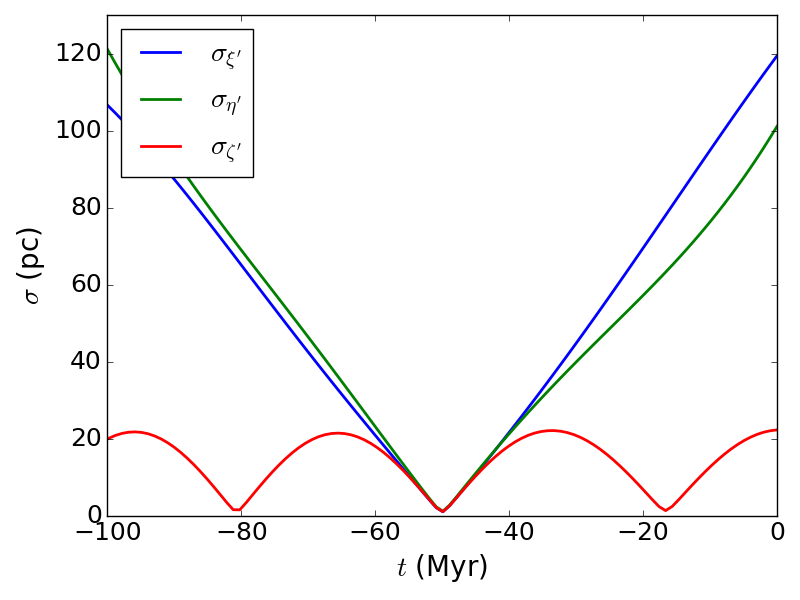}
\includegraphics[scale=0.35]{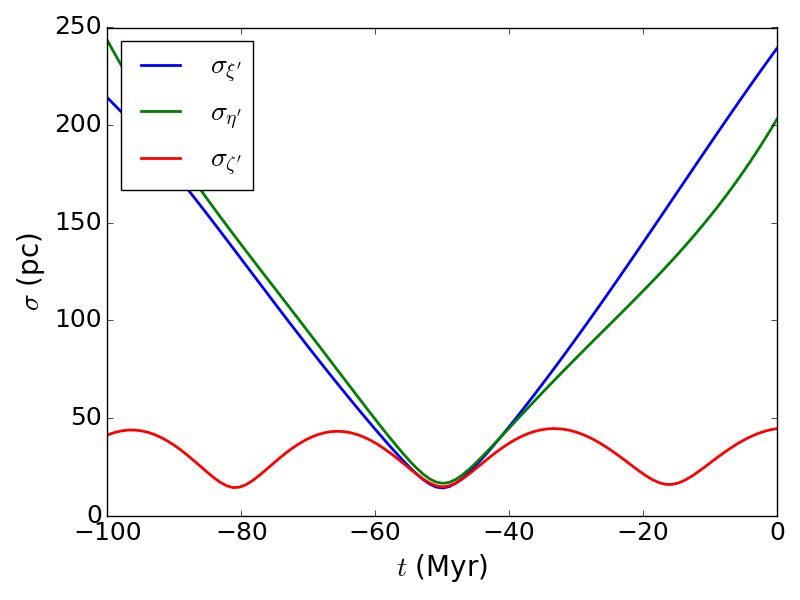}
\caption{Dispersion in three positional components (radial $\xi^\prime$, azimuthal $\eta^\prime$, and vertical $\zeta^\prime$) obtained from the integration back in time of simulated YLAs following orbit-1 (Sect.~\ref{subsec:generation-YLA}). We show two different initial conditions: IC-1 (top) and IC-2 (bottom). The  scales are different in the two panels. }
\label{fig:simple-case}
\end{figure}

Here we present the results of the orbital integration of the simulated YLA with orbit-1 (Sect.~\ref{subsec:generation-YLA}). In this simulation, we did not take into account observational errors. In Figure~\ref{fig:simple-case} we show the dispersion in positions as a function of time in the three coordinates (radial, azimuthal, and vertical). We note that in the absence of observational errors and when the true Galactic potential is used, we can clearly see that the time of formation (50~Myr) presents a minimum dispersion in positions, and therefore we can recover the age of the simulated YLA. We reach  equivalent conclusions when dealing with older associations (e.g. 200 Myr). 

In the vertical direction, the dispersion oscillates with a period\footnote{This period is about half of the period of the epicyclic frequencies of the individual orbits of the members of the simulated YLA. This is expected according to the epicyclic approximation, e.g. as described by the equations of Appendix A in \citet{Asiain99}.} of $\sim 35$~Myr. This value is comparable to the ages of the YLAs and, consequently, there are times of minimum dispersion that can be confused with the formation time. Thus, the coordinate $\zeta^{\prime}$ is not useful for a determination of the dynamical age.

The dispersions in the radial and azimuthal directions oscillate, but with a dominant cycle of longer period ($\sim150$~Myr). The dispersion in position in the azimuthal direction has, additionally, a secular evolution (not noticed here due to the short integration time, but see Eq.~\ref{eq:AsiainEta} and text in Sect.~\ref{subsec:dispersion}). This secular evolution of $\sigma_{\eta^{\prime}}$ makes it possible to distinguish between the times of minimum dispersion caused by the epicyclic oscillations and the minimum related to the birth time of the YLA because the latter must be an absolute minimum.  We note that a higher initial dispersion in positions and velocities (Fig.~\ref{fig:simple-case}, bottom) does not change the location of the minimum dispersion, but  makes it less peaked. To conclude, from now on we focus our trace-back analysis on the azimuthal component in positions, $\eta^{\prime}$, to study the dynamical age of the YLAs.

\subsection{Treatment of observational errors}
\label{subsec:dispersion}

In the previous section we  computed the dispersion in positions through the time integration of the orbits. From now on, we  refer to this dispersion with no observational errors taken into account as the intrinsic dispersion ($\sigma_\textup{int}$). In reality, the dispersion in the present time includes the effect of the observational errors. Thus, by integrating the orbits back in time we can only determine the propagated observed dispersion ($\sigma_\textup{obs}$). At a given time $t$ of our orbital trace-back analysis, the two dispersions (intrinsic and observed) are related through
\begin{equation}
\sigma_\textup{obs}^2 = \sigma_\textup{int}^2 + \sigma_\textup{err}^2
\label{eq:intrinsic-disp}
,\end{equation}
where $\sigma_\textup{err}$ is the dispersion due to the observational errors of all the members of the association. This is a dispersion term that arises from the effect of the observational errors in the present time. 

To estimate the effect of the observational errors at any time of the trace-back analysis, we used the analytical expressions derived by \citet{Asiain99} which are based on the epicyclic approximation. These equations predict the dispersion in positions and velocities as a function of time, given an initial dispersion. Let $\vec{x^0}$= ($\xi^\prime_0$, $\eta^\prime_0$, $\zeta^\prime_0$, $\dot{\xi}^\prime_0$, $\dot{\eta}^\prime_0$, $\dot{\zeta}^\prime_0$) be the vector of present ($t=0$) position and velocity of each star in a YLA and [$\sigma_{\vec{x^0_j}}, \ \textup{j}=1,6$] the current dispersions. Then, the dispersion in the azimuthal position component at any time $t$ is 
\begin{eqnarray}
\label{eq:AsiainEta}
\sigma^2_{\eta^\prime}& = & \left[ \displaystyle\frac{2 \omega_\odot \mbox{A}}{\mbox{B}} \left( t - \displaystyle\frac{\sin (\kappa t)}{\kappa} \right) \right]^2 \sigma^2_{\xi^\prime_0} +
      \sigma^2_{\eta^\prime_0} + \nonumber \\
      & & + \left[ \displaystyle\frac{2 \omega_\odot}{\kappa^2} \left( \cos (\kappa t) - 1  \right) \right]^2 \sigma^2_{\dot{\xi}^\prime_0} + \nonumber \\
      & & + \left[ \displaystyle\frac{1}{\mbox{B}} \left( \mbox{A} t - \displaystyle\frac{\omega_\odot}{\kappa} \sin (\kappa t) \right) \right]^2 \sigma^2_{\dot{\eta}^\prime_0}~, \nonumber\\
\end{eqnarray}
where A and B are the Oort constants and $\kappa$ is the epicyclic frequency. The equations for the other components can be found in  Appendix A (Eq.~A.2) of \citet{Asiain99}. As pointed out by the authors, the dispersion in each component oscillates around a constant value except for $\sigma_{\eta^\prime}$, whose average value increases with time. More in detail, as can be seen in Eq. \ref{eq:AsiainEta}, the dispersion in the azimuthal positions has two secular terms: one is proportional to the initial dispersion in the radial positions $(\sigma_{\xi^\prime_0})$ and the other to the initial dispersion in the azimuthal velocity $(\sigma_{\dot{\eta}^\prime_0})$. 

We used this epicyclic approximation only to estimate the impact of the observational errors at each time step of the orbital integration. The initial dispersions due to the observational errors [$\sigma_{\textup{err}, \ \vec{x^0_j}}, \ \textup{j}=1,6$] are an estimation of the dispersion due to the observational errors in the present. To compute this, we first converted the individual astrometric errors (in parallax and proper motions) and spectroscopic errors (radial velocities) of each star to curvilinear errors and then computed the median error in each coordinate. After estimating $\sigma_\textup{err} (t=0)$ and computing $\sigma_\textup{err}$ as a function of the backwards time with Eq. \ref{eq:AsiainEta}, we can derive an estimation of the intrinsic dispersion with time as $\sigma_\textup{est}^2 =\sigma_\textup{obs}^2 -\sigma_\textup{err}^2$. 

We emphasise that we only used the epicyclic approximation to estimate the effect of the errors as a function of time. In contrast, we computed the intrinsic and the propagated observed dispersion through full orbital integrations, which is more precise than Eq. \ref{eq:AsiainEta} based on the epicyclic approximation.

\subsection{Impact of the observational errors}
\label{subsec:error-prop}

\begin{figure}
\centering
\includegraphics[scale=0.35]{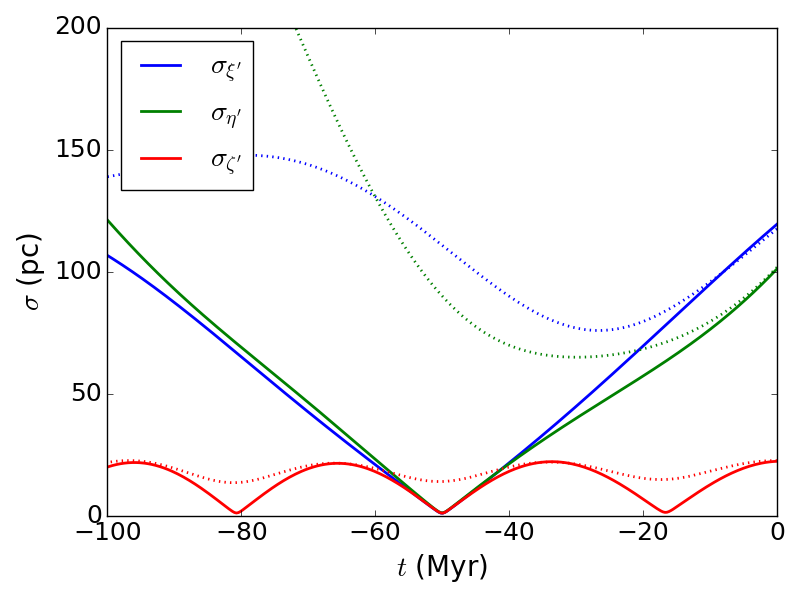}
\caption{Intrinsic dispersion (solid lines) and propagated observed dispersion (dotted lines) in positions from the integration back in time of a simulated YLA with initial conditions IC-1 and following orbit-1. }
\label{fig:effect-errors}
\end{figure}

\begin{figure*}
 \centering
 \includegraphics[scale=0.33]{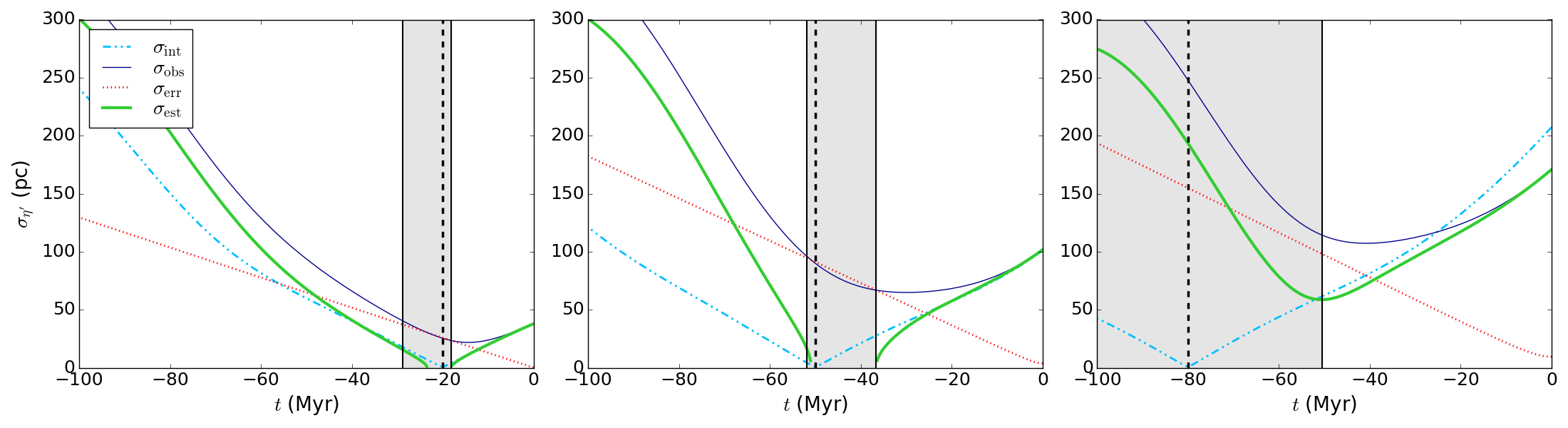}
 \caption{Evolution in time of the dispersion in the $\eta^\prime$ coordinate for a simulated YLA with initial conditions IC-1 and following orbit-1, from left to right, with ages of $20$, $50$ and $80$ Myr (marked with a vertical black dashed line). The different colour lines are the intrinsic dispersion ($\sigma_\textup{int}$, cyan dot-dashed line), the propagated observed dispersion ($\sigma_\textup{obs}$, dark-blue solid line), the dispersion due to observational errors ($\sigma_\textup{err}$, red dotted line) and the estimated intrinsic dispersion ($\sigma_\textup{est}$, green thick solid line). The grey-shaded region represents the uncertainty in the dynamical age determined. }
 \label{fig:determination-DA-axi}
\end{figure*}

Here  we evaluate the effects of the observational errors in the derivation of the dynamical age of the YLAs with the trace-back analysis. In Figure~\ref{fig:effect-errors} we show the quantities $\sigma_\textup{int}$ (solid lines) and $\sigma_\textup{obs}$ (dotted lines) obtained in our trace-back analysis for the simulated YLA with initial conditions IC-1 and following orbit-1 (see Sect.~\ref{subsec:generation-YLA}). We see how, from the beginning, the propagated observed dispersions start to differ from the intrinsic dispersions. This figure shows the extent of the impact that the observational errors have on the estimation of the dynamical age. Indeed, the radial and azimuthal components present a minimum at an earlier time than the simulated age of the association. Furthermore, the minimum dispersion is higher than the simulated value at birth. 

Similarly, in Fig.~\ref{fig:determination-DA-axi} we present the dispersions obtained when we integrate back in time simulated YLAs with different ages. In this case we subtract the dispersion due to observational errors (red dotted line) from the propagated observed dispersion (dark blue solid line) as explained in Sect.~\ref{subsec:dispersion} to obtain the estimated intrinsic dispersion (green thick solid line). The  estimated intrinsic dispersion $\sigma_\textup{est}$ resembles the {true} intrinsic dispersion (cyan dot-dashed line) for short integration times.  
In some cases (e.g. Fig.~\ref{fig:determination-DA-axi}, left and middle panels), the dispersion due to observational errors is larger than the propagated observed dispersion ($\sigma_\textup{err}>\sigma_\textup{obs}$), meaning that the formulas for the error propagation (Sect.~\ref{subsec:dispersion}) are overestimating the errors. Even so, the approximation allows us to determine, not a time of minimum dispersion but a plausible range for the dynamical age, as we see in Sect.~\ref{subsec:criteria}. In general, for the cases studied here, with approximately circular orbits and relatively young ages, we find the approximation to be a good one and to render unbiased determinations of the dynamical age. 

For the association of $80$~Myr (Fig.~\ref{fig:determination-DA-axi} right panel), we see that the observed dispersion in the present (which includes the effect of observational errors) appears to be slightly smaller than the intrinsic dispersion. This is due to the cut in distance at 500~pc (see Sect.~\ref{subsec:generation-YLA}). This cut eliminates some members that, before introducing the observational errors, were {close} to the Sun, but {far} from it after error convolution. In other words, the number of members needed to compute the propagated observed dispersion $\sigma_\textup{obs}$ is smaller than those needed to compute the intrinsic dispersion $\sigma_\textup{int}$. However, this has little effect on our determination of the dynamical age, as we see in Sect.~\ref{subsec:criteria}.

\subsection{Criteria to determine the dynamical age}
\label{subsec:criteria}

After the analysis of previous sections, we describe here the criteria that we used to determine the dynamical age of the YLAs and its associated uncertainty. The dynamical age is defined as the time in which the association is more concentrated in the positional space. In practice, we computed it as the time of minimum dispersion in the estimated intrinsic dispersion $\sigma_\textup{est}$ determined as explained in Sect.~\ref{subsec:dispersion} and illustrated in Fig.~\ref{fig:determination-DA-axi} (thick green solid line). 

We adopted an observational threshold of $15$~pc for the birth size of an association. This is the size of the nearest star forming regions \citep{Blaauw91}. Only if the dispersion measured at the time of minimum dispersion is below this threshold do we consider that the association has a size compatible with that of a forming region. If the threshold is not reached, we cannot  provide a dynamical age and such associations will need a deeper study. We note that this is a strong assumption and that this threshold is based on the current knowledge of the star formation conditions and thus should be revised in the future. 

The criteria that we follow to determine the dynamical age are as follows:
\begin{enumerate}
\item In the most favourable cases, we find a minimum  $\sigma_\textup{est}$ which is below the observational threshold. Then, we report the time of this minimum as the dynamical age of the association. To determine the uncertainty, we follow the next approach. First, we note that the birth of a YLA cannot occur while the dispersion decreases with backwards time. Thus, the effective lower uncertainty is zero (in other words, the  lower limit coincides with our determination of the age itself). 
Second, we take as upper limit the time at which the association reaches the size of the observational threshold of 15~pc. 
\item When the minimum  $\sigma_\textup{est}$ cannot be found because the dispersion due to observational errors ($\sigma_\textup{err}$) is larger than the propagated observed dispersion ($\sigma_\textup{obs}$), we cannot provide a single value for the dynamical age. Instead, we provide a range equivalent to the uncertainty range defined in criterion 1. This case is illustrated in the left and middle panels of Fig.~\ref{fig:determination-DA-axi} (grey  shaded area). 
\item If we find a minimum  $\sigma_\textup{est}$ that is higher than the observational threshold (e.g. minimum at about 55 Myr in Fig.~\ref{fig:determination-DA-axi} right panel), we provide only a lower limit for the dynamical age. In this case the curve tells us only that $\sigma_\textup{est}$ decreases  to this time, and therefore that the association cannot be younger than this. Beyond this age, the dispersion increases solely dominated by the errors, and thus we do not have information of the intrinsic dispersion. 
\item In the most unfavourable cases, we find that $\sigma_\textup{est}$ strongly increases in the recent past and it might not present a minimum. As is seen in Sect.~\ref{subsec:DA}, this is the case of two real YLAs. Such cases need a deeper evaluation (see discussion in Sect.~\ref{sec:discussion}). 
\end{enumerate}

We admit that our criterion of  15 pc is arbitrary. However, despite the limitations of our method and its assumptions, we find that it gives better age determinations and uncertainties than other methods. We explore this in Appendix \ref{App:A} where we estimate the dynamical age by determining the minimum dispersion and its error by performing Monte Carlo resampling of the observational errors (i.e. without going through all of the steps in our methodology). We find that, in this way, the dynamical ages are biased and the uncertainties are underestimated. Our previous tests with simulations with our full methodology and also the tests presented in Sect.~\ref{subsec:restricted-sample} show that our methodology gives, with limitations for large ages and a small number of members, correct estimations of the dynamical ages and their derived uncertainties.

\subsection{Impact of restricted samples}
\label{subsec:restricted-sample}

\begin{figure*}
 \centering
 \includegraphics[scale=0.33]{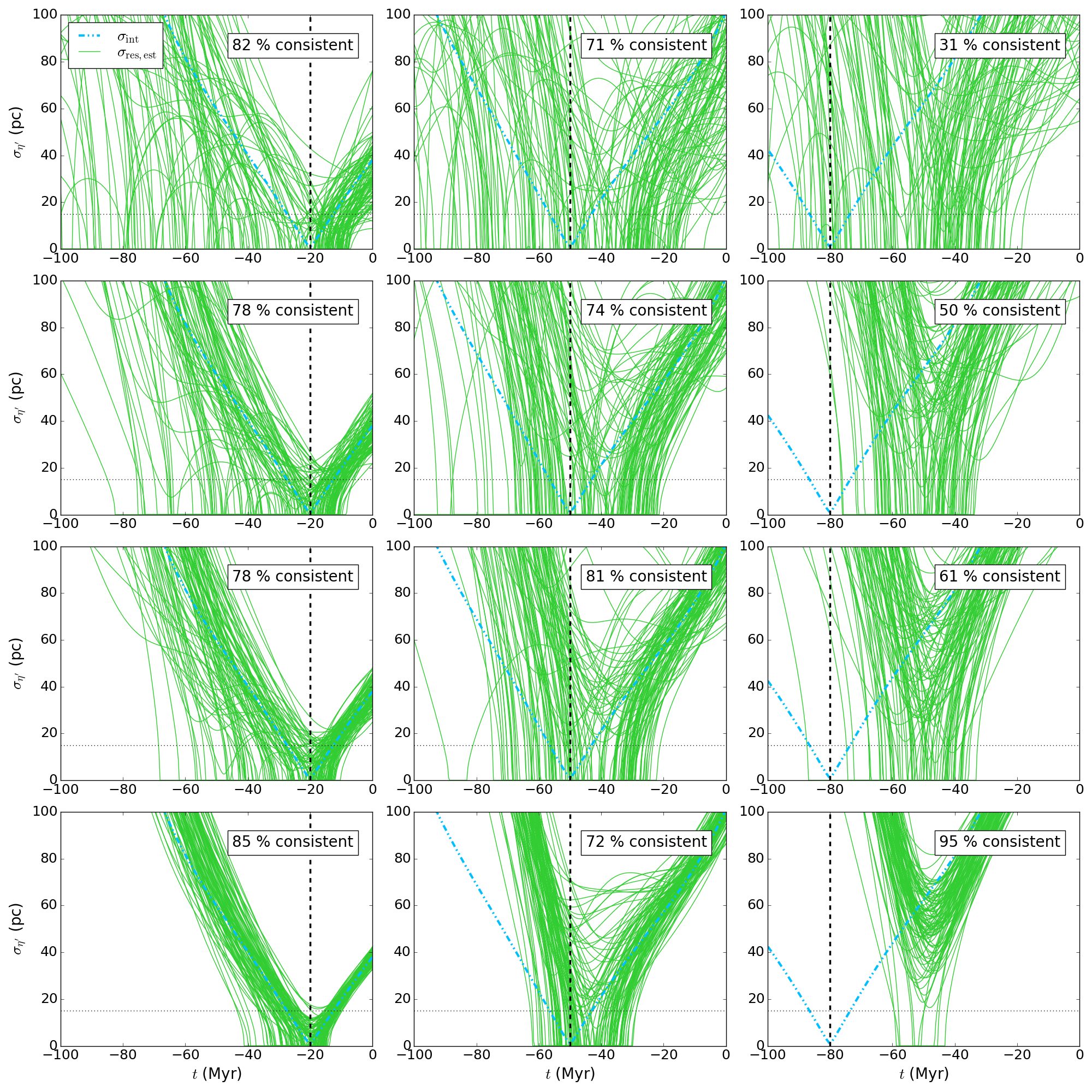}
 \caption{Evolution in time of the dispersion in the $\eta^\prime$ coordinate for a simulated YLA using a resampling technique where we only consider (from top to bottom) 5, 15, 25, and 100 members of the 500 members of the association. The initial conditions are IC-1 with orbit-1 and ages (from left to right) of 20, 50, and 80~Myr (marked with a vertical black dashed line). The different  lines are the intrinsic dispersion ($\sigma_\textup{int}$, cyan dot-dashed line) and the estimated intrinsic dispersion of each of the 100 trials ($\sigma_\textup{res,est}$, green solid lines). In each panel, we indicate the percentage of success in determining the dynamical age.}
\label{fig:restricted-sample}
\end{figure*}

\begin{table}
\centering
\begin{tabular}{|l|c||ccc|}
\hline
\hline
Memb. &  Crit.   & 20~Myr & 50~Myr & 80~Myr  \\
\hline
\hline
    & 1   &  7  & 0   & 4   \\
5   & 2   & 92  & 91  & 79  \\
    & 3   &  1  & 9   & 17  \\  
\hline
    & 1   & 20  & 2   & 3   \\
15  & 2   & 72  & 76  & 47  \\
    & 3   & 8   & 22  & 50  \\  
\hline
    & 1   & 23  &  1  & 3   \\
25  & 2   & 68  & 69  & 37  \\
    & 3   & 9   & 30  & 60  \\  
\hline
    & 1   & 38  & 4   & 0   \\
100 & 2   & 60  & 62  & 5   \\
    & 3   & 2   & 34  & 95  \\  
\hline
\hline
\end{tabular}
\caption{Percentage of cases in which we have applied a given criteria (defined in Sect.~\ref{subsec:criteria}) for each of the panels in Fig.~\ref{fig:restricted-sample}.} 
\label{tab:rest-samp}
\end{table}

Before we applied our methodology to the observational sample, we tested the impact of having data for a restricted number of members of the association. We performed a  Jackknife-like resampling, considering only a subset of 5, 15, 25, and 100 members (from top to bottom in Fig.~\ref{fig:restricted-sample}) for an association of 20, 50, and 80~Myr (from left to right). We repeated this process 100 times for each subset of members, we computed the estimated intrinsic dispersion (green solid lines) by subtracting the dispersion due to observational errors to the propagated observed dispersion (see Sect.~\ref{subsec:error-prop}), and we determined the dynamical age and its uncertainty following our methodology. 

In Table \ref{tab:rest-samp} we show, for each of the 12 panels in Fig.~\ref{fig:restricted-sample}, the percentage of cases in which we have applied a given criteria (Sect.~\ref{subsec:criteria}). The total percentage of success in determining the dynamical age, in other words the number of times that our age determination is consistent with the {true} age of the association within the uncertainty, is shown in each of the panels of Fig.~\ref{fig:restricted-sample}. 

We see that in most cases we have used criterion 2 and we do not find any case under criterion 4. This means that in the majority of the cases we cannot provide a given age and its uncertainty but a range for the dynamical age. 

As expected, when the association is old (four rightmost panels), the number of successful cases is low, except when the number of members is large (100 members, bottom right panel). With this we show that our methodology combined with the current data uncertainties is not suitable for old associations. We also note that in the case of old associations our method would not be suitable due to the uncertainties on the galactic potential. For the rest of the panels, we see that in general the fraction of successful cases increases with decreasing age and increasing number of members, as expected. We see that the fraction of cases for which we are able to provide a given age and its uncertainty range (criterion 1) instead of only a range (criterion 2) increases in the same direction, especially for young ages (leftmost panels). For the bottom middle panel, the unexpected decrease in successful cases might be attributed to the fact that we only use 100 cases for this estimation; the numbers are very similar to the panel above it, and we note that the number of cases under criterion 1 has increased.

For  the panels with ages younger than 80 Myr (three first columns), the fraction of success is between 70\% and 90\%. If the errors were Gaussian, we would expect a fraction of 68\% of successful cases. This means that our uncertainty determination is correct and yields an error equivalent to between the 1$\sigma$ and 2$\sigma$ fraction in a Gaussian case, i.e. between 68\% and 95\%. Finally, we see that the fraction of success is high even for a small number of members (five members, first column of panels), but that in these cases only a range of plausibility is given for the dynamical age (criterion 2) and that in about 20--30\% of the cases this range does not include the true age but is close to it. Our results for a low number of members, therefore, must be taken with caution.

%__________________________________________________________________

\section{Results for the known YLAs using TGAS}
\label{sec:results}

In this section we use the data of our set of observed YLAs to determine, first, their dynamical ages using the methodology explained in Sect.~\ref{sec:method-sim} and, second, their place of birth.

\subsection{Dynamical ages}
\label{subsec:DA}

\begin{figure*}
\centering
\includegraphics[scale=0.33]{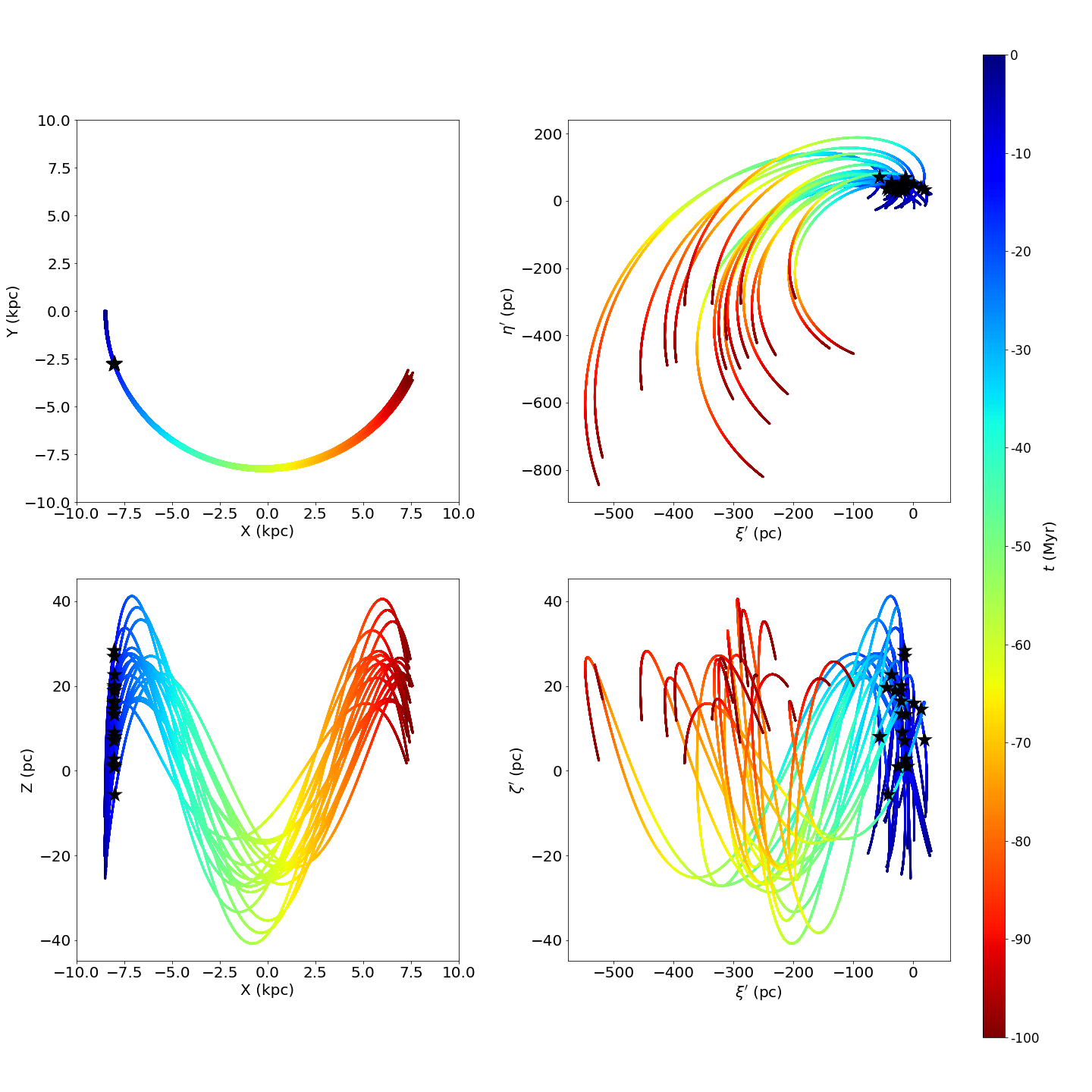}
\caption{Orbital projections of the bona fide members of the $\beta$~Pic association, integrated under the axisymmetric potential. We show the $(X,Y)$ projection (top left), the $(X,Z)$ projection (bottom left), the $(\xi^\prime,\eta^\prime)$ projection (top right), and the $(\xi^\prime, \zeta^\prime)$ projection (bottom right). For each of the members, we mark with a star the position at which we obtain the minimum dispersion.}
\label{fig:2D-bpic}
\end{figure*}

\begin{figure*}
\centering
\includegraphics[scale=0.36]{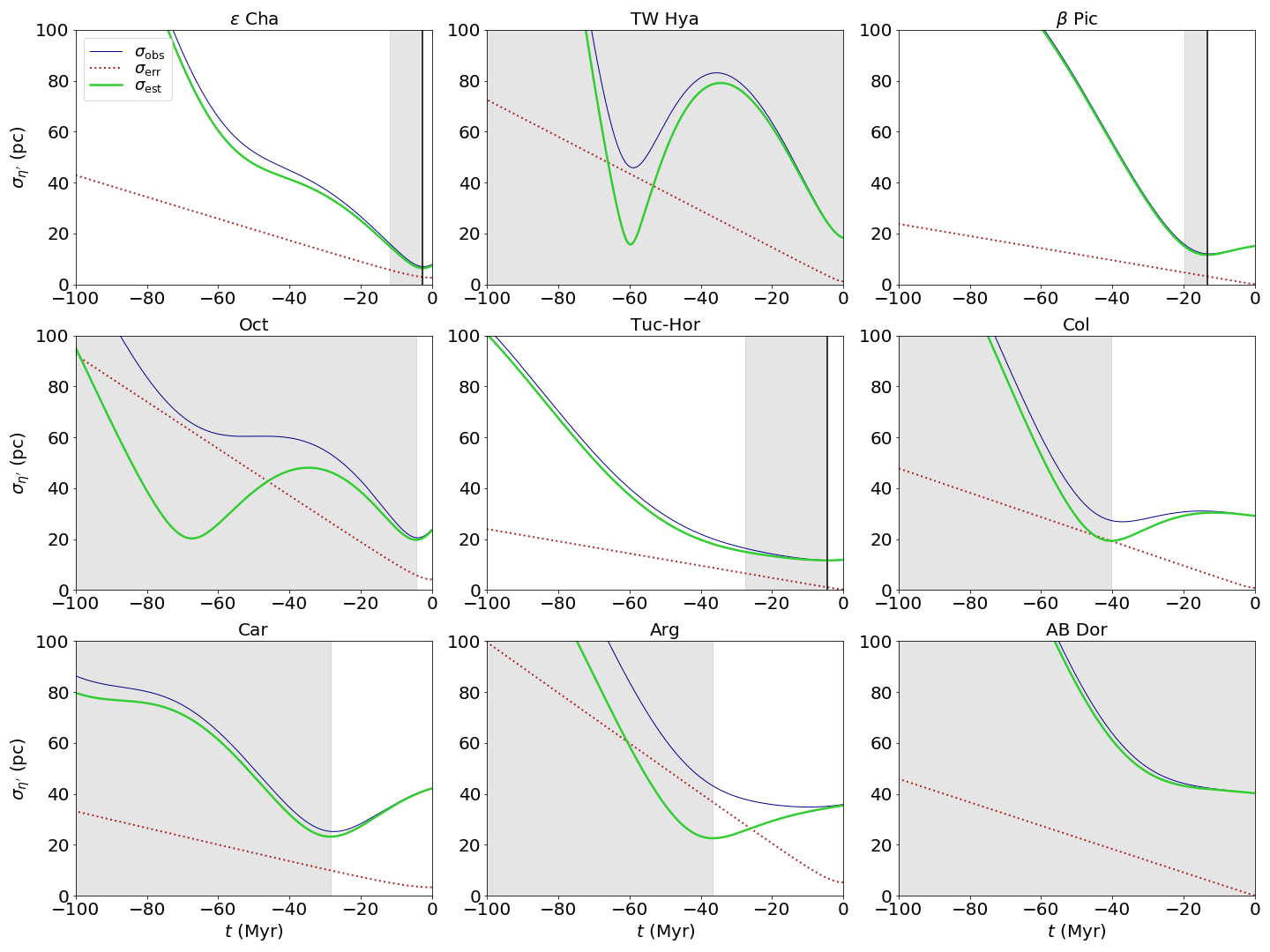}
\caption{Evolution with time of the dispersions in $\eta^\prime$ coordinate for the observed associations presented in Table~\ref{tab:YLAs-review}. We show the  propagated observed $\sigma_\textup{obs}$ (dark blue solid lines), dispersion due to observational errors $\sigma_\textup{err}$ (red dotted lines), and estimated intrinsic dispersion $\sigma_\textup{est}$ (green thick solid lines). The grey shaded areas represent the region of uncertainty of the dynamical age. }
\label{fig:DA-observed-YLAs}
\end{figure*}

We integrated back in time the orbits of the bona fide members with full 6D phase-space data of each YLA in Table~\ref{tab:YLAs-review}, after having applied a $2\sigma$ clipping in the velocity space (see Sect.~\ref{sec:sample} for further details on the sample selection). We computed the dispersions due to observational errors $\sigma_\textup{err}$ in the present time as the median error in the coordinates. We used only the axisymmetric potential since the non-axisymmetries do not have a significant effect on the young associations considered here (see Sect.~\ref{subsec:spiral-arms} for a quantification of the effect of non-axisymmetric structures, such as  spiral arms).

As an example, in Figure~\ref{fig:2D-bpic} we present the orbits of the bona fide $\beta$~Pic members. The left panels show the projections in galactocentric coordinates ($X,Y$) and ($X,Z$) (top and bottom, respectively), while the right panels show the projections in curvilinear heliocentric coordinates $(\xi^\prime,\eta^\prime)$ and $(\xi^\prime, \zeta^\prime)$ (top and bottom, respectively). As expected, the orbits are nearly circular around the Galactic centre and the vertical trajectories describe a harmonic oscillation. For each of the members, we mark with a star the position at which we obtain the minimum dispersion (see below). We note that it is in the curvilinear coordinate system where we most clearly visualise the grouping of the orbits during the first $\sim30$~Myr.

In Table~\ref{tab:YLAs-DA} we report the dynamical ages of the YLAs obtained, and in Fig.~\ref{fig:DA-observed-YLAs} we show the dispersions as function of time for each association. The $\epsilon$~Cha, $\beta$~Pic, and Tuc-Hor associations are the  optimal cases: we find a minimum in the estimated intrinsic dispersion ($\sigma_\textup{est}$), which is smaller than the observational threshold (criterion 1 in Sect.~\ref{subsec:criteria}). This results in a dynamical age determination of $3^{+9}_{-0}$~Myr, $13^{+7}_{-0}$~Myr, and $5^{+23}_{-0}$~Myr, respectively.  

The Col, Car, and Arg associations only present one minimum in the estimated intrinsic dispersion curve. However, this minimum is larger than the observational threshold and therefore, according to criterion 3, we only report a lower limit in the dynamical age. 

The Oct association is a special case where we find two minima in the estimated intrinsic dispersion curve, neither of which has a dispersion smaller than the observational threshold. In this case, we report a lower limit for the dynamical age according to criterion 3. As pointed out by \citet{Riedel17}, Oct is one of the most distant YLAs for which the authors found it difficult to differentiate its members from field stars. 

For the TW Hya and the AB Dor associations we find inconclusive results. In both cases, the dispersion increases in the recent past. We note that these are the most adverse cases since one of them presents the smallest number of bona fide members (only seven) and the other has the oldest age (50--150~Myr). See Sect.~\ref{subsec:restricted-sample} for details on the influence of the number of members on the dynamical age determination and Sect.~\ref{sec:discussion} for further discussion.

\begin{table}
\centering
\begin{tabular}{lcccc}
\hline
\hline
Association    & Crit.     & Dynamical age  & Literature age \\
               &           & (Myr)          & (Myr)          \\
\hline
$\epsilon$ Cha & 1         & 3$^{+9}_{-0}$  & $5-8$          \\
TW Hya         & 4         & $-$            & $3-15$         \\
$\beta$ Pic    & 1         & 13$^{+7}_{-0}$ & $10-24$        \\
Oct            & 3         & $>$4           & $20-40$        \\
Tuc-Hor        & 1         & 5$^{+23}_{-0}$ & $30-45$        \\
Col            & 3         & $>$40          & $30-42$        \\
Car            & 3         & $>$28          & $30-45$        \\
Arg            & 3         & $>$37          & $35-50$        \\
AB Dor         & 4         & $-$            & $50-150$       \\
\hline
\hline
\end{tabular}
\caption{Results on the dynamical ages of YLAs considered in this study determined with our orbital trace-back analysis. We indicate in each case which  criterion is used (see Sect.~\ref{subsec:criteria}). We also present the ages compiled by \citet{Riedel17} from different sources using methods such as spectroscopic determinations and isochrone fitting. }
\label{tab:YLAs-DA}
\end{table}

\subsection{Place of birth}
\label{subsec:place}

\begin{figure*}
\begin{center}
\includegraphics[scale=0.38]{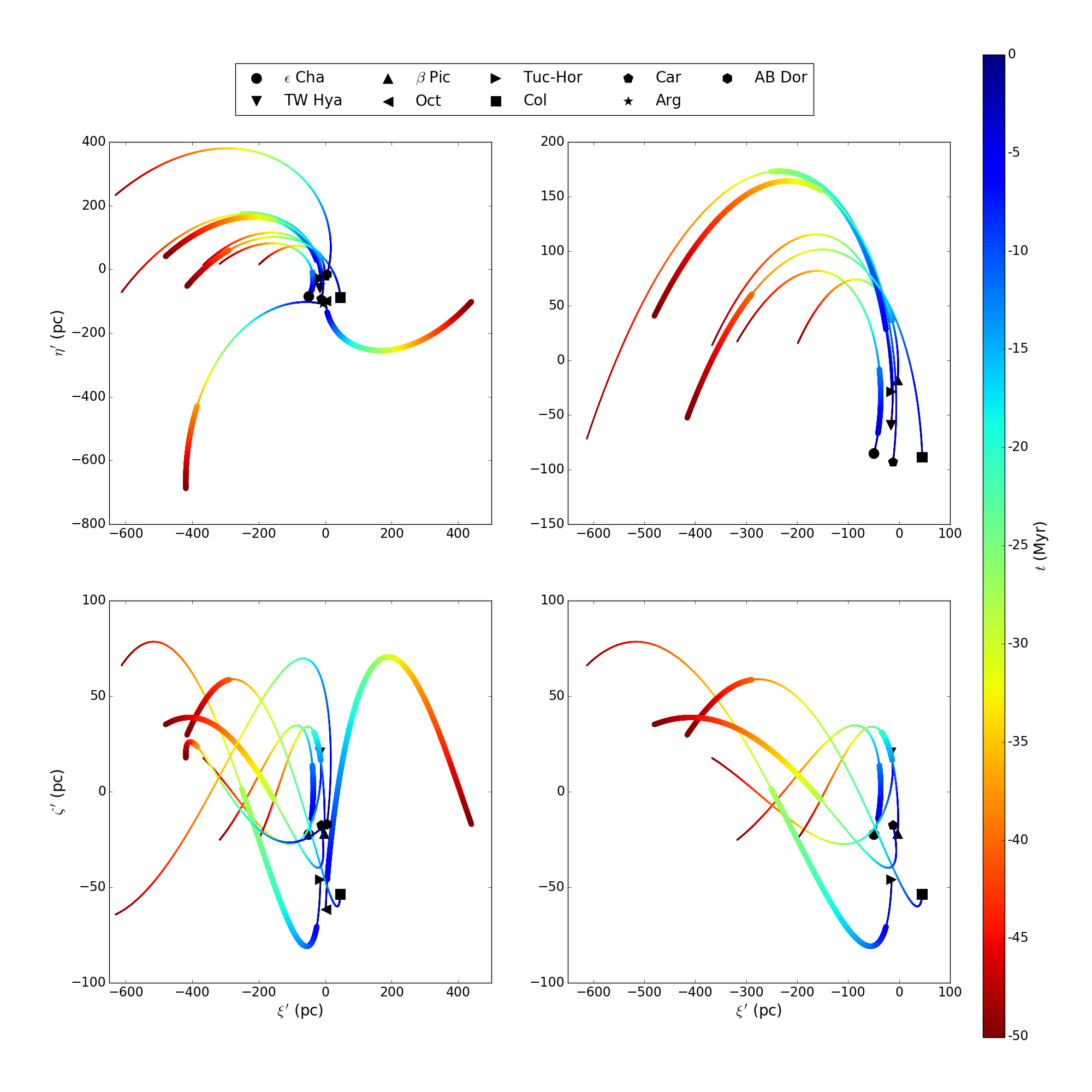}
\caption{Curvilinear heliocentric $(\xi^\prime,\eta^\prime)$ projection (top) and $(\xi^\prime,\zeta^\prime)$ (bottom), colour-coded as a function of time, of the orbits of the centroids of the associations considered in this study (Table \ref{tab:YLAs-review}). The panels on the right are a zoom of the panels on the left, removing the Octans, Argus, and AB Dor associations (see text). The larger dots mark the region where the associations are formed according to our dynamical age determination ranges from Table~\ref{tab:YLAs-DA}.}
\label{fig:place-birth}
\end{center}
\end{figure*}

The orbital analysis of the YLAs is a tool not only for determining the dynamical age of these systems, but also for studying the region of formation. In this section we consider the centroid of each of the nine YLAs studied (see Table~\ref{tab:YLAs-review}) and integrate their orbits back in time using the axisymmetric potential. Figure~\ref{fig:place-birth} shows the resultant orbits in curvilinear heliocentric coordinates both in the plane $(\xi^\prime,\eta^\prime)$ (top panels) and the vertical projection $(\xi^\prime,\zeta^\prime)$ (bottom panels), colour-coded as a function of time. We note that in the present, all the associations are spread over a region of $90\times140\times80$~pc and most of them are in the fourth Galactic quadrant (see also Table~\ref{tab:YLAs-review}). 

We also see that the heliocentric velocity components of the YLAs are very similar except for the Octans, Argus, and AB Dor associations. Without considering these three associations (zoom shown in the panels on the right), the mean heliocentric velocity of the rest of the YLAs is $(\bar{U},\bar{V},\bar{W})=(-11,-20,-6)$~km~s$^{-1}$, with a dispersion of $2-3$~km~s$^{-1}$ in each component. This similar kinematics means that they have very similar orbits, which suggests that they may have co-evolved or that they are related in some way. 

In Figure~\ref{fig:place-birth} we have also included the information regarding the age of these associations: the thicker lines indicate the region where the association is formed according to our dynamical age determination ranges (Table~\ref{tab:YLAs-DA}).  We see that Col and Car have similar ages ($\sim 30-40$~Myr) and at the time of their birth they were nearby, in a region of about $100$~pc, slightly above the Galactic plane ($\zeta^\prime\sim$ 40~pc). $\epsilon$~Cha and $\beta$~Pic are younger, with ages of 3 and 13~Myr, respectively, and at the time of their birth were separated by roughly 50~pc. On the other hand, our age determination for Tuc-Hor is a bit uncertain and it shows that it could have been formed in any of the two groups. However, according to the spectroscopic ages found in the literature, it is more plausible that it was formed together with the Col and Car associations. We could not derive a dynamical age for TW Hya. However, if we consider the determinations found in the literature, its position at the time of formation coincides with the youngest formation region mentioned.

In short, our results seem to indicate that there were two regions of formation: one about $30-40$~Myr ago that gave birth to the Tuc-Hor, Col, and Car associations, and the other $3-20$~Myr ago that gave birth to $\epsilon$~Cha, TW Hya, and $\beta$~Pic. We note, however, that the vertical motion of the TW Hya association is in the opposite direction to the other two tentatively co-forming associations. 
The age differences  of the associations suggest that they were not formed together, but instead in different star bursts.

%__________________________________________________________________

\section{Discussion}
\label{sec:discussion}

Here  we present a thorough discussion of the results obtained in the previous sections. First, we compare the results obtained in Sect.~\ref{sec:results} with studies in the literature. Second, we provide possible explanations for the cases showing discrepancies or inconclusive results. At this point, we review the future perspectives for the study of the YLAs when more accurate data will be available. Specifically, we evaluate the foreseen accuracies for the Second \textit{Gaia} Data Release (DR2). Finally, we quantify how the use of non-axisymmetric potentials including spiral arms becomes more important when older associations are considered.

\subsection{Comparison with other studies}
\label{subsec:lite-comp}

Most of our determinations of the dynamical ages are consistent with the ages from the literature, either spectroscopic or isochronal, within the uncertainties. Only for the Tuc-Hor association does our determination yield a younger age than the spectroscopic values.

Furthermore, most of our dynamical age determinations are similar to other dynamical ages found in the literature, where  TW~Hya and $\beta$~Pic are the most studied associations.
\citet{delaReza06} performed a trace-back analysis of TW~Hya similar to ours, but based on \textit{Hipparcos} astrometry. With only four members, they determined a dynamical age of $8.3\pm0.8$~Myr. Instead, we are not able to determine a dynamical age for this association with our sample of seven stars (TWA~1, TWA~5, TWA~6, TWA~9, TWA~19, TWA~21, and TWA~25), only with two members in common with their study (TWA~1 and TWA~19). However, we find that the dynamical age of this association is very dependent on the sample selection: if we consider only TWA~5, TWA~6, TWA~21, and TWA~25 we find an age of $3^{+6}_{-0}$~Myr, which is consistent with \citet{delaReza06}. Again, we emphasise that the results based on small samples should be considered with care. To estimate the error in the dynamical age, \citet{delaReza06} used Monte Carlo realisations to simulate the errors in velocity space (1~km s$^{-1}$). However, they did not take into account the uncertainties on the parallax.

More recently, \citet{Weinberger13} performed a trace-back analysis of TW~Hya with a larger (19 stars; TWA~1, TWA~5, TWA~21, and TWA~25 in common with our study) but more heterogeneous sample (parallaxes from \textit{Hipparcos} and from their own survey, and proper motions from UCAC3, \citealt{Zacharias09}). Their analysis resulted in a minimum dispersion 2~Myr ago, but they argued that this result was not significant because the dispersion decreased by only a very small amount. However, they did not take into account the propagation of the observational errors or the effects of neglecting the Galactic potential since they used linear trajectories.

Later, \citet{Ducourant14} using their own membership criterion for TW~Hya (based on the convergent point and refined by the trace-back analysis), and their own parallaxes and proper motions from UCAC4 \citep{Zacharias12}, found a dynamical age of $7.5\pm0.7$~Myr. They started from a sample of 25 stars, 5 of which are in common with our study (TWA~1, TWA~5, TWA~6, TWA~9, and TWA~21), but eventually they  discarded 9 stars (only TWA~1 remained in common with our study) arguing that they systematically drifted from the mean. Here we have used a statistically more robust criterion to remove outlier stars (Sect.~\ref{sec:sample}). They study the uncertainty on the dynamical age coming from the sample contamination by means of a Jackknife-like resampling where they eliminate $20\%$ of the sample from each run. However, they did not consider the error introduced by the observational uncertainties of each member or the effect of neglecting the Galactic potential since they use linear propagation to trace-back the positions of the stars. \citet{Donaldson16}, using the same data sample, repeated the analysis but included the effect of observational errors via a Monte Carlo sampling and obtained a completely different result of $3.8\pm1.1$~Myr. This shows the impact that observational errors have on dynamical age determinations.

Regarding the dynamical age of the $\beta$~Pic association, the study of \citet{Mamajek14} uses the members reported in \citet{Zuckerman04} together with \textit{Hipparcos} parallaxes, proper motions either from \textit{Hipparcos} or UCAC4, and radial velocities from a compilation of sources, and taking only the members with better precisions. We share five members (HIP~10680, HIP~10679, HIP~88399, HIP~95270, and HIP~84586). They explore three different methodologies for the trace-back orbits, but they could not find a conclusive dynamical age. At this point, we reinforce the idea that to succeed with a trace-back analysis it is mandatory to have extremely accurate data and membership.

For the TW Hya, Oct, and AB Dor associations we found ambiguous dynamical ages (i.e. several dispersion minima or a dispersion always increasing with backwards time), and thus they demand a deeper study. We note that our study of TW~Hya is based on few members, which makes the analysis more uncertain. The AB~Dor association is the largest one in terms of number of members, and even though previous authors have reported ages of $\sim 50$~Myr \citep{Close05}, it is now  believed to be as old as $100-150$~Myr \citep{Luhman05,Bell15}. This age poses a challenge for the dynamical trace-back analysis. A future membership analysis combined with accurate astrometric data from \textit{Gaia} should clarify these inconsistencies. In general, all the dynamical ages presented here are a preliminary result obtained with the first \textit{Gaia} data and should be updated with further releases. This  especially applies to the cases where the analysis is based on fewer than 20 members, i.e. $\epsilon$~Cha, TW~Hya, Oct, and Car.

The scenario that we have proposed in which several associations might have been formed together is also supported by other studies. \citet{Elmegreen93} proposed a three-burst scenario. A first generation of star formation began about $60$~Myr ago when the local gas was compressed due to the passage of the Carina spiral arm, which led to the formation of the Lindblad ring. The age of this burst is close to the dynamical ages of one of our suggested groups (Tuc-Hor, Col, and Car). A second generation of stars occurred when the Lindblad ring fragmented $20$~Myr ago, giving birth to structures such as Orion, Sco-Cen, or Perseus. According to our determinations, the age of $\beta$~Pic fits, within the uncertainties,  this second burst. Finally, a third generation of stars was formed more recently in the regions of Ophiuchus and Taurus;  according to our results, these regions could have given birth to $\epsilon$~Cha. \citet{deZeeuw99} and \citet{Sartori03} found younger ages for the Sco-Cen complex ($5-15$~Myr) which are compatible with our dynamical ages of $\epsilon$~Cha and $\beta$~Pic. Finally, \citet{Mamajek00}, \citet{Ortega02}, and \citet{Fernandez08} also related the formation of these two associations to the Sco-Cen complex, in support of our findings.

\subsection{Caveats}
\label{subsec:caveats}

In this section we review our assumptions and analyse which are the most critical points  responsible for the inconclusive results found in some associations.

First of all, our trace-back method relies on a certain membership classification. Although the orbital analysis allows us to identify contaminants, the trace-back analysis is hindered if there is membership misclassification. The majority of the membership algorithms are based on the kinematic properties of the association,  but since different YLAs occupy similar positions in the velocity space, this classification is complex and currently still under study \citep{Riedel17,Elliott16}. 

It is also possible that we are considering several associations together or that not all the stars in a given YLA  were formed at the same time but rather in different star bursts \citep{Barenfeld13}. A detailed analysis of the orbits using future \textit{Gaia} data with more precise astrometry and spectroscopy will allow us to find possible subgroupings. Binarity can also play a role in the results we obtain. If there are unresolved binaries in our sample, this would introduce an error in the radial velocity of the star. As we prove with our simulations, accurate radial velocity is essential for a successful orbital analysis.

For the modelling we have assumed the epicyclic approximation to propagate the observational errors back in time. While this is a good approximation for small integration times and quasi-circular orbits,  the vast majority of cases in our analysis, we have also seen that in some cases it may not be accurate enough. Therefore, this will deserve reassessment in future analysis with older YLAs and YLAs with more eccentric orbits, which  might be discovered in the coming years.

\subsection{Perspectives with \textit{Gaia} DR2}
\label{subsec:DR2}

\begin{figure*}
 \centering
 \includegraphics[scale=0.33]{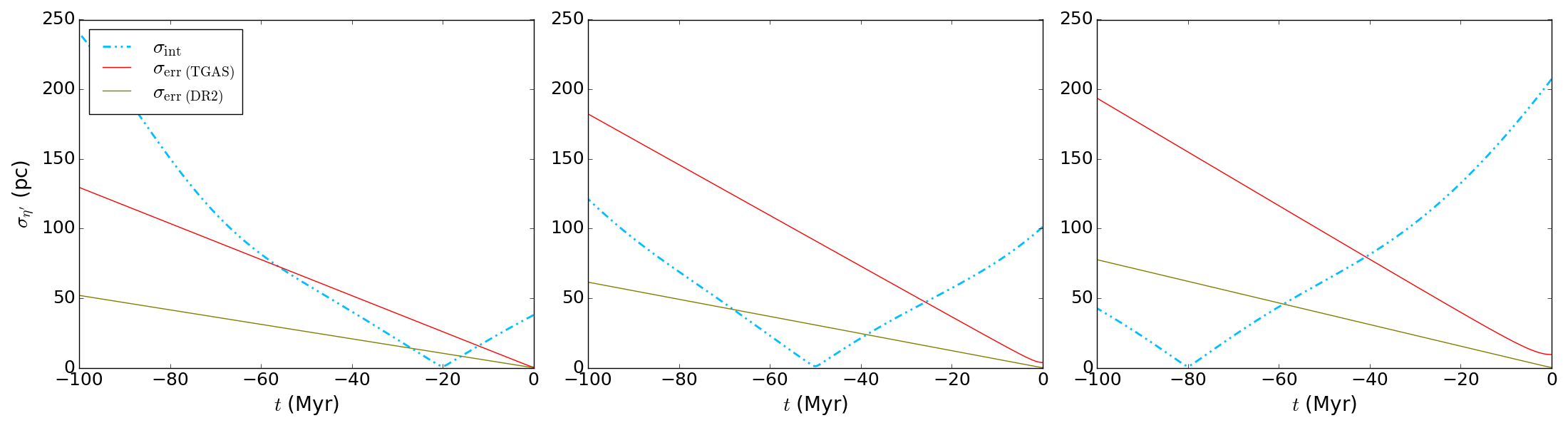}
 \caption{Evolution of the intrinsic dispersion (cyan dot-dashed lines) and the dispersion due to observational errors in the $\eta^\prime$ coordinate for a simulated YLA following orbit-1 of $20$, $50,$ and $80$ Myr (from left to right) with initial conditions IC-1. Two different data scenarios are considered, namely TGAS (red solid lines) and DR2 (olive solid lines).}
 \label{fig:errors-prop}
\end{figure*}

The \textit{Gaia} DR2 will be published\footnote{https://www.cosmos.esa.int/web/gaia/dr2} on  25 April 2018 and it will include the five-parameter solution for sources up to magnitude $G=21$ and radial velocities for stars brighter than $G\sim12$. This will constitute an excellent catalogue to revisit the study of the YLAs. In this section, we aim to quantify the improvement in the determination of the dynamical age of the YLAs that \textit{Gaia} DR2 will entail. 

To simulate the DR2 scenario we used the uncertainties from \textit{Gaia} DR2 and $\sigma_\textup{RV} =1$~km~s$^{-1}$ expected from the Gaia-RVS spectroscopy\footnote{https://www.cosmos.esa.int/web/gaia/science-performance}. In addition, we needed an assumption for the $(V-I)$ colour distribution apart from that of the visual magnitude $V$ (Sect.~\ref{subsec:generation-YLA}). We took it again as a Gaussian centred at $(V-I)=0.9$ with a dispersion of 0.7~mag, consistent with our data sample (Sect.~\ref{sec:sample}). We note that the colour is only a second-order effect in the computation of the astrometric uncertainties. 
 
In  Figure~\ref{fig:errors-prop} we compare the propagated errors $\sigma_\textup{err}$ of our current data set (TGAS) calculated as explained in Sect.~\ref{subsec:generation-YLA} with the expected ones for DR2. We do it for a set of simulated YLAs with ages of $20$, $50,$ and $80$~Myr, respectively.  We also show the intrinsic dispersion curves. The time at which the two curves intersect corresponds to the time when the propagated observed dispersion is composed in equal parts of the intrinsic dispersion and that due to the effect of the errors. We note that $\sigma_\textup{err}$ depends on both the data scenario considered and on the age of the simulated YLA. Old YLAs are indeed currently more dispersed in space, and thus farther away from the Sun, meaning that its median observational errors are larger.

We observe that in the DR2 scenario, the orbit of a young association of $20$~Myr (left panel) can be integrated almost until birth without impact of the errors, i.e. $\sigma_\textup{err}<\sigma_\textup{int}$. 
For an intermediate-age YLA (middle panel) the dispersion is dominated by the errors only after about $40$ Myr instead of $25$ Myr with TGAS data. For the oldest association (right panel) this happens for $\sim60$~Myr with \textit{Gaia} DR2, but for $\sim40$~Myr in the TGAS case.

To conclude, the improvement of DR2 with respect to the TGAS data is substantial. We see that with the TGAS scenario the errors already become important  at about half of the age of the YLAs for intermediate-age and old associations. Instead, with DR2 the backwards integration is valid, i.e. not dominated by the errors, much closer to the age of the association. We also note that in this analysis we have not yet subtracted the model for the errors, in which case the possibility of getting closer to the real age would be even better.

\subsection{Effect of spiral arms}
\label{subsec:spiral-arms}

\begin{table}
\centering
\tabcolsep=0.45cm
\begin{tabular}{c|cc|cc}
\hline
\hline 
 & \multicolumn{2}{|c|}{ orbit-1 }  & \multicolumn{2}{c}{orbit-2} \\[0.2mm]
\hline
Age   & $\Delta$ Age & $\epsilon$ Age & $\Delta$ Age  & $\epsilon$ Age \\
(Myr) & (Myr)        & $(\%)$         & (Myr)         &  $(\%)$        \\[0.2mm]
\hline
20  &   0.04 &  0.2 &  0.04 &   0.2 \\
50  &   2.5  &  5   &  4    &   7   \\
80  &  11    & 14   & 16    &  20   \\
200 &  20    & 10   & 33    &  16   \\[0.2mm]
\hline
\hline
\end{tabular}
\caption{Absolute and relative errors in the determination of the dynamical age of a YLA when neglecting the effect of the spiral arms. We consider two types of YLAs, one with an almost circular orbit (orbit-1, left column) and one with a more eccentric orbit (orbit-2, right column).}
\label{tab:effect-spiral-arms}
\end{table}

\begin{figure}
\centering
\includegraphics[scale=0.35]{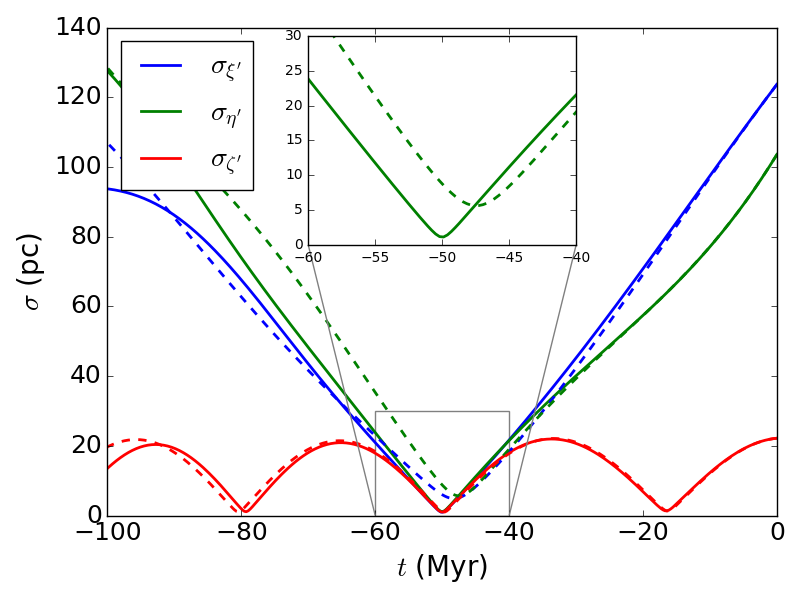}
\caption{Dispersion in positions from the integration back in time of the simulated YLA generated with initial conditions IC-1 and following orbit-1 (Sect.~\ref{subsec:generation-YLA}). The true evolution is computed with a potential including spiral arms (solid lines) and is  compared with the evolution computed using the axisymmetric model (dashed lines). } 
\label{fig:effect-spiral-arms}
\end{figure}

Most of the previous work of trace-back analysis in the literature has neglected the effects of the non-axisymmetric perturbations on the orbits of the YLAs (see Sect.~\ref{sec:introduction}). In this section, we use a set of simulated YLAs to quantify for the first time the effects of the spiral arm perturbation on the derivation of the dynamical age.  For the YLAs that we study here, we see that these effects might be neglected, but they will become a relevant matter when associations of longer lifetimes are studied with the \textit{Gaia} catalogues in the near future.

We use  a non-axisymmetric potential which accounts for the spiral arms in addition to the axisymmetric potential described in Sect.~\ref{subsec:integration}. 
The 3D spiral model consists of the PERLAS spiral arms from \citet{Pichardo03}. The locus has a pitch angle of $15.5\degree$ and a shape that was estimated by \citet{Drimmel01}. We take a pattern speed of $\Omega=21$~km~s$^{-1}$~kpc$^{-1}$ and a mass of $0.04\%$ of the disc mass. These values are in agreement with the values proposed in \citet{Antoja11}.

We consider the two centroids defined in Sect.~\ref{subsec:generation-YLA}: the one with a circular orbit (orbit-1, similar to AB Dor) and the one with a slightly more eccentric orbit (orbit-2, similar to Hyades). We also explore YLAs of different ages, namely 20, 50, 80, and 200 Myr, generated following Sect.~\ref{subsec:generation-YLA}. In Figure~\ref{fig:effect-spiral-arms} we show the time evolution of the dispersion in positions for the circular orbit case (orbit-1) with an age of $50$~Myr.  In particular, we superpose the true time evolution (with the spiral arm potential, solid lines) and the evolution assuming the axisymmetric model (dashed lines). We observe that although the spiral arms are neglected, the time of minimum dispersion for this relatively young association agrees with the simulated age of the YLA.

The comparison between the simulated dynamical age of all our associations and the age obtained when we neglect the effects of the spiral arms  is presented in Table~\ref{tab:effect-spiral-arms}. We find that when dealing with young associations of less than $20$~Myr, for both orbit-1 and orbit-2 the axisymmetric potential is a good approximation: the error introduced is smaller than $\sim 1$~Myr. For older associations, this error increases with age reaching values up to $20$~Myr for an association of $200$~Myr with a quasi-circular orbit. The errors are larger for orbit-2. We note that the relative error seems to settle to values of 15--20\% for the two oldest cases. We have checked that these conclusions also hold  for the different values of the initial dispersion  proposed in Sect.~\ref{subsec:generation-YLA}. 

From this analysis and the results presented in Sect.~\ref{subsec:error-prop} and \ref{subsec:DR2}, we see that the small differences in the determination of the dynamical age between using the axisymmetric potential or a spiral arm potential are smaller than the error induced by the current observational errors for the cases presented here. Therefore, the effect of the spiral arm potential can be neglected. Nonetheless, when the observational data is more accurate and we consider older associations (with longer integration times), the effect of the spiral arms should not be overlooked. 

%__________________________________________________________________

\section{Conclusions}
\label{sec:conclusions}

In this work we determined the dynamical age and place of birth of the youngest and most populated YLAs. This is the first time that the dynamical age and its associated uncertainty have been derived in a uniform way for such a large group of YLAs and using the recent \textit{Gaia} data.

Our method consists in integrating the orbits of the YLA members back in time and finding the time of minimum dispersion in the azimuthal component of the positions. We  have tested and fine-tuned the method using mock YLAs. In particular, we see that the observational errors of the combination of astrometry from TGAS and current complementary on-ground radial velocities have a major impact on the determination of dynamical ages. For this, we have successfully modelled the effects of the errors and  taken them into account to estimate the dynamical age. In addition, we prove here with simulations that for the youngest YLAs with nearly circular orbits and the current observational errors, the axysimmetric potential is a sufficient model to integrate their orbits without biasing the results. Thus, we can neglect effects of the non-axisymmetries. The simulations presented with mock YLAs also show that all the results obtained with few members should be taken with extreme care.

With the current data we are able to determine or constrain the dynamical age for seven out of the nine YLAs studied, namely $\epsilon$ Chamaeleontis, $\beta$ Pictoris, Octans, Tucana-Horologium, Columba, Carina, and Argus. Moreover, in most of the cases our results are compatible with spectroscopic and isochrone fitting ages  and with other dynamical determinations elsewhere. Our determinations  thus offer an independent, uniform, and valuable constraint on the ages of YLAs in the solar neighbourhood. 

For the controversial cases of TW~Hya and AB Dor, which have very few members or are old and might be related to two different burst of formation, more accurate data is necessary to reach a plausible age determination. Further data releases of the \textit{Gaia} mission and other surveys will increase the list of members of the YLAs and improve the quality of the current data.  We have quantified that the next data release of \textit{Gaia}, DR2, which will supply the first all-sky set of accurate and homogeneous radial velocities, indeed offers notable improvement in the determination of the dynamical ages, even for older associations. At this point, we also emphasise the need of spectroscopic surveys of radial velocities for faint YLA members with accuracies comparable to that of \textit{Gaia} DR2 tangential velocities. For older associations or those following more eccentric orbits, the effects of the spiral arms will become important, but this can be accounted for in a straightforward way with our method.

We see that currently all but three of  the YLAs in our set  have very similar orbits. This can either suggest that most of the associations were formed from the same molecular cloud or that our relatively small selection of YLAs is biased towards these positions. Very interestingly, we see hints of two different regions and epochs of star formation that could have given birth to two different groups of YLAs comprising Tuc-Hor, Col, and Car (30--40 Myr ago) and $\epsilon$~Cha, TW Hya, and $\beta$~Pic (3--15 Myr ago). Additionally, our orbital analysis would be compatible with the first generation of YLAs being related to the passage of the Carina spiral arm and the second one to the formation of the Sco-Cen complex. Future \textit{Gaia} data will provide a more complete sample of YLAs and a more comprehensive  scenario describing when and where these associations were formed, and their relation with known Galactic structures of star formation.

\begin{acknowledgements}\\
We thank the anonymous referee for his/her useful comments that helped improving the manuscript.
This work was supported by the MINECO (Spanish Ministry of Economy) through grant ESP2016-80079-C2-1-R (MINECO/FEDER, UE) and ESP2014-55996-C2-1-R (MINECO/FEDER, UE) and MDM-2014-0369 of ICCUB (Unidad de Excelencia 'María de Maeztu').
This work has made use of data from the European Space Agency (ESA) mission {\it Gaia} (\url{https://www.cosmos.esa.int/gaia}), processed by the {\it Gaia} Data Processing and Analysis Consortium (DPAC, \url{https://www.cosmos.esa.int/web/gaia/dpac/consortium}). Funding for the DPAC has been provided by national institutions, in particular the institutions participating in the {\it Gaia} Multilateral Agreement.
\end{acknowledgements}

%-------------------------------------------------------------------

\bibliographystyle{aa}
\bibliography{refs}

\appendix

\section{Monte Carlo errors}
\label{App:A}

\begin{figure*}
 \centering
 \includegraphics[scale=0.33]{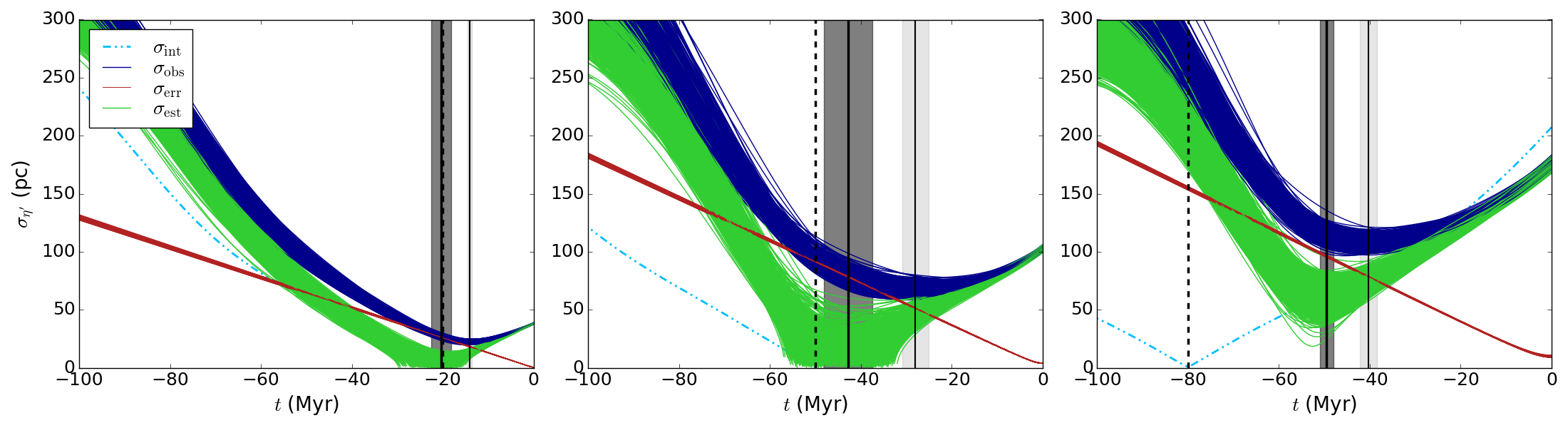}
 \caption{Evolution in time of the dispersion in the $\eta^\prime$ coordinate for a simulated YLA with initial conditions IC-1 and following orbit-1 (from left to right) with ages of $20$, $50,$ and $80$ Myr (marked with a vertical black dashed line) with 1000 Monte Carlo realisations of the observational errors. The different  lines are the intrinsic dispersion ($\sigma_\textup{int}$, cyan dot-dashed line), the propagated observed dispersion for each Monte Carlo realisation ($\sigma_\textup{obs}$, dark blue solid lines), the dispersion due to observational errors ($\sigma_\textup{err}$, red lines), and the estimated intrinsic dispersion ($\sigma_\textup{est}$, green solid lines). The light grey shaded region represents the uncertainty in the dynamical age determined with the propagated observed dispersion, and the dark grey shaded region represents the uncertainty in the dynamical age determined with the estimated dispersion. }
 \label{fig:MC-errors}
\end{figure*}

In this appendix we describe the  1000 Monte Carlo realisations of the observational errors of associations with different ages. For each realisation we computed the propagated observed dispersion (dark blue solid lines in Fig.~\ref{fig:MC-errors}). As a first approximation, we did not perform our full methodology, which corrects for the observational errors effect, but only took into account the errors through the Monte Carlo sampling. We therefore used each of these dark blue curves to determine a dynamical age of the association by looking for the time of minimum dispersion, and we computed the mean and the standard deviation of all realisations (vertical solid black line and light grey area, respectively). We see that with this approximation, the dynamical ages are systematically underestimated and the error bars do not include the true age of the association. Several studies have applied a similar methodology to estimate the uncertainties in the dynamical age \citep[e.g.][]{delaReza06, Weinberger13, Donaldson16}. However, here with these tests, we see the need to include a more advanced treatment of the observational errors.

In order to better take into account the observational errors, we follow the procedure explained in Sect.~\ref{subsec:error-prop} and subtract the dispersion due to errors (red lines) to the propagated observed dispersion to obtain the estimated intrinsic dispersion (green solid lines). Then we determinate the dynamical ages with the estimated intrinsic curves. The resultant dynamical age is indicated as a vertical thick solid line and the corresponding uncertainty as a dark grey area. In all cases, considering the analytical estimation of the observational errors improves the determination of the dynamical age. In consequence, we believe it is essential to include the analytical treatment of the observational errors for the success of the dynamical age determination. 

In addition, we see that the Monte Carlo uncertainties are still greatly underestimated  since they do not include the true age of the association, especially for intermediate-age and old associations. That is why in Sect.~\ref{subsec:criteria} we propose different criteria to estimate the uncertainties. Although they are slightly arbitrary criteria, which  should be revised in the future when more is known about the initial conditions of star formation, we find them necessary to provide accurate uncertainties. When we compare the error bars in Figs.~\ref{fig:determination-DA-axi} (our methodology) and \ref{fig:MC-errors} (Monte Carlo errors), we see that our methodology performs better and that our error bars include the true value of the age, while  the Monte Carlo ones do not.

\end{document}